\documentclass[10pt,conference]{IEEEtran} 
\IEEEoverridecommandlockouts
\usepackage{cite}
\usepackage{amsmath,amssymb,amsfonts}
\usepackage{algorithmic}
\usepackage{graphicx}
\usepackage{textcomp}
\usepackage{xcolor}
\def\BibTeX{{\rm B\kern-.05em{\sc i\kern-.025em b}\kern-.08em
    T\kern-.1667em\lower.7ex\hbox{E}\kern-.125emX}}
\usepackage{amsmath,bm}
\usepackage{comment}
\usepackage{enumitem}
\usepackage{siunitx} 
\usepackage{multirow}
\usepackage{booktabs}
\usepackage{soul}
\usepackage{pifont}
\usepackage{float} 
\usepackage{subfigure}
\usepackage{wrapfig}

\usepackage[bookmarks=true,breaklinks=true,colorlinks,linkcolor=blue,citecolor=blue,urlcolor=blue]{hyperref}
\usepackage{xspace}

\newcommand{\anx}{\anx\xspace}

\newcommand{\Anx}{\Anx\xspace}

\newcommand{\ansx}{\ansx\xspace}

\newcommand{\design}{JustQ\xspace}

\begin{document}
\date{}

\title{\design: Automated Deployment of Fair and Accurate Quantum Neural Networks\vspace{-10pt}}

\author{Ruhan Wang$^1$, Fahiz Baba-Yara$^2$, Fan Chen$^1$\\
\textit{$^1$Luddy School of Informatics, Computing, and Engineering},
\textit{$^2$Kelley School of Business}\\
\textit{Indiana University, Bloomington, IN, USA} \\
E-mail: \{ruhwang, fababa, fc7\}@iu.edu\\
\vspace{-38pt}
}

\maketitle

\begin{abstract}
Despite the success of Quantum Neural Networks (QNNs) in decision-making systems, their fairness remains unexplored, as the focus primarily lies on accuracy. This work conducts a design space exploration, unveiling QNN unfairness, and highlighting the significant influence of QNN deployment and quantum noise on accuracy and fairness.
To effectively navigate the vast QNN deployment design space, we propose \design, a framework for deploying fair and accurate QNNs on NISQ computers. It includes a complete NISQ error model, reinforcement learning-based deployment, and a flexible optimization objective incorporating both fairness and accuracy. Experimental results show~\design outperforms previous methods, achieving superior accuracy and fairness. This work pioneers fair QNN design on NISQ computers, paving the way for future investigations.
\end{abstract}


\vspace{-2pt}
\section{Introduction}
\label{sec:intro}
\vspace{-2pt}

\textbf{Motivation}.
Quantum Neural Networks (QNNs)\cite{Sukin2019_vqcc14, chu2022qmlp, PatelST22vqcdate, wang2022vqcdac, guan2022verifying} have become powerful tools for automated decision-making, spanning finance\cite{focardi2020quantum}, healthcare~\cite{parsons2011possible}, and drug discovery~\cite{cao2018potential}.
While classical machine learning models are known to exhibit susceptibility to social biases~\cite{dwork2012fairness}, current works on QNNs predominantly emphasize accuracy~\cite{Sukin2019_vqcc14, chu2022qmlp, PatelST22vqcdate, wang2022vqcdac} with limited discussion on fairness~\cite{guan2022verifying}. 
We summarize related works in Table~\ref{tab:related_work} and identify three challenges. 
\underline{\textit{First}}, 
none of the prior work performed a holistic evaluation of the accuracy and fairness of QNNs. 
\underline{\textit{Second}}, 
existing work lacks a comprehensive method for modeling and measuring quantum noises in Noisy Intermediate-Scale Quantum (NISQ) computers~\cite{preskill2018quantum}, despite their significant impact on accuracy and fairness.
\underline{\textit{Third}}, 
prior efforts solely focused on noise-aware training~\cite{chu2022qmlp, PatelST22vqcdate, wang2022vqcdac}. However, our preliminary studies show that accuracy gains from training can be compromised by imprudent and often unknown synthesis settings during the deployment phase.

\textbf{Contributions}.
This work introduces a novel framework, ~\design, which addresses fairness and accuracy in QNNs simultaneously. Our contributions are as follows:
\begin{itemize}[leftmargin=*, topsep=0pt, partopsep=0pt, itemsep=0pt]
    \item We explored the design space using IBM quantum computers and found that: (1) QNN deployment has a greater impact on accuracy and fairness than training, and (2) QNN accuracy and fairness are closely linked during deployment, highlighting the need for automated exploration methods.
    \item We present the~\design~framework with three components: (1) an NISQ error model and measurement method; (2) a reinforcement learning-based approach for automated, fair, and accurate QNN deployment; and (3) a flexible optimization objective combining fairness and accuracy in a customizable reward function for diverse application goals.    
    \item We evaluate~\design~on various QNNs and show that~\design surpasses prior deployment methods in both accuracy and fairness. Furthermore, \design's adaptable optimization objective empowers users to tailor it to specific application needs.    
\end{itemize}

\begin{table}[t!]\setlength{\tabcolsep}{1.5pt}
\begin{center}
\caption{A summary of related works.}
\vspace{-8pt}
\label{tab:related_work} 
\footnotesize
\begin{tabular}{|l|c|c|c|c|c|c|}\toprule
\multirow{2}{*}{\textbf{Scheme}} & \multicolumn{2}{c|}{\textbf{Metric}} &\multicolumn{2}{c|}{\textbf{Error Model}} &\multicolumn{2}{c|}{\textbf{Design Phase}} \\ \cline{2-7} 
  &\textbf{Acc.}  &\textbf{Fairness}  &\textbf{Real\_Device} &\textbf{Meas\_Method} &\textbf{Train.}   &\textbf{Deploy.}\\\midrule  
C14~\cite{Sukin2019_vqcc14}  &\multirow{4}{*}{\textbf{\ding{52}}}  &\multirow{4}{*}{\textbf{\ding{56}}} &\text{\ding{56}} &\text{\ding{56}} &\multirow{5}{*}{\textbf{\ding{52}}} &\multirow{5}{*}{\textbf{\ding{56}}} \\ \cline{1-1}\cline{4-5}
QMLP~\cite{chu2022qmlp}  & & &\text{\ding{52}}  &\text{\ding{56}} & & \\\cline{1-1}\cline{4-5} 
DATE22~\cite{PatelST22vqcdate}  & & &\text{\ding{52}}  &\text{\ding{56}}  &  & \\\cline{1-1}\cline{4-5}
DAC22~\cite{wang2022vqcdac}  & & &\text{\ding{52}}  &\text{\ding{56}}  &  &  \\\cline{1-3}\cline{4-5} 
CAV22~\cite{guan2022verifying}    & \textbf{\ding{56}} & \textbf{\ding{52}} &\text{\ding{56}}  &\text{\ding{56}}  &  & \\\hline
\textbf{\design} & \text{\ding{52}} &\text{\ding{52}} &\text{\ding{52}}  &\text{\ding{52}}  &\text{\ding{56}} &\text{\ding{52}} \\\bottomrule 
\end{tabular}
\end{center}
\vspace{-26pt}
\end{table}

\textbf{Limitations and Scope}.
QNNs have demonstrated advantages over their classical counterparts in theory~\cite{riste2017demonstration, Lloyd2018lett} and small-scale practical implementations~\cite{genois2021quantum, hu2019quantum}. 
In this work, we initiate research on fair and accurate QNNs using IBM quantum computers, paving the way for future investigations in this domain. 
Machine learning fairness is an evolving research area lacking a widely agreed-upon definition. We adopt the concept of \textit{individual} fairness~\cite{dwork2012fairness}, which advocates treating similar individuals equitably for similar outcomes.
The limitation on the implementation scale (up to 27 qubits) is due to current NISQ technology capabilities. As technology continues to advance~\cite{preskill2018quantum}, the implementation scale and performance enhancement of~\design will also improve. 

\section{Preliminary}
\label{sec:back}

\subsection{NISQ Hardware}
\vspace{-2pt}
NISQ noises~\cite{nachman2020unfolding, nielsen2002quantum, ahsan2022quantum} lead to \textit{readout}, \textit{gate}, and \textit{crosstalk} errors.
\textit{Readout} errors refer to incorrect qubit measurement, e.g., reading $|1\rangle$ while the qubit is in the $|0\rangle$ state and vice versa. They are modeled as a 2$\times$2 probability matrix and mitigated with readout-error correction~\cite{nachman2020unfolding}.
\textit{Gate} errors are dominated by 2-qubit gates and modeled as depolarizing noise~\cite{nielsen2002quantum}: Err($\rho$) = (1-$p$)$\rho$ + $p$$\frac{I}{2^n}$, where $p$ is the system error rate depending on NISQ computers and the circuit. Err($\rho$), $\rho$, and $n$ refer to the noise channel, the noiseless density matrix, and the qubit count, respectively. 
\textit{Crosstalk} errors result from non-ideal qubit interactions in systems with over 10 qubits, causing deviations from the ideal behavior of quantum gates and circuits. The occurrence and impact of \textit{crosstalk} are influenced by the total number and spatial arrangement of 2-qubit gates within the circuit~\cite{ahsan2022quantum}.

\noindent
\underline{\textbf{Limitations in Prior Models}}. 
Prior noise-aware QNNs~\cite{chu2022qmlp, PatelST22vqcdate, wang2022vqcdac} focused on \textit{gate} errors but ignored \textit{crosstalk} errors. Recent work~\cite{ahsan2022quantum} shows that \textit{crosstalk} can cause a 20\% increase in phase-flip error and a 33\% decrease in gate coherence in a small 9-qubit circuit on the \texttt{IBM\_Melbourne} computer.
Moreover, none of the previous work~\cite{chu2022qmlp, PatelST22vqcdate, wang2022vqcdac} provides details on error estimation and measurement, which are time-consuming and can introduce significant latency overhead.

\vspace{-2pt}
\subsection{Quantum Neural Networks}
\vspace{-2pt}

\begin{figure}[t!]
\centering
\includegraphics[width=1\linewidth]{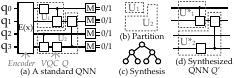}
\vspace{-20pt}
\caption{A standard QNN and its approximate synthesis.}
\label{f:background}
\vspace{-20pt}
\end{figure}

QNNs~\cite{Sukin2019_vqcc14, chu2022qmlp, PatelST22vqcdate, wang2022vqcdac} are a sequence of \textit{quantum gates} that operate on a set \textit{qubits}. As Figure~\ref{f:background}(a) shows, a QNN comprises a data encoder, $E(x)$, embedding a classical input $\mathbf{x}$ into a quantum state $|\mathbf{x}\rangle$, a Variational Quantum Circuit (VQC), $\mathbf{Q}$, generating the output state, and a measurement layer, $M$, mapping the output quantum state to a classical vector.

\textbf{QNN Fairness}.
We consider \textit{individual} fairness~\cite{guan2022verifying} for equitable treatment of similar inputs, ensuring unbiased outcomes. 
A QNN is \textit{fair} if and only if no ($\epsilon$, $\delta$)-$bias$ pairs exist in the dataset  $\mathcal{T}$. Such a pair is characterized by the input state distance between $E(x_i)$ and $E(x_j)$ being within a threshold $\epsilon$, while the difference between their output $\mathbf{Q}(E(x_i))$ and $\mathbf{Q}(E(x_j))$ exceeds a threshold $\delta$, formulated as:
\vspace{-2pt}
\begin{equation}
\left[\mathbf{D}(E(x_i), E(x_j))\le\epsilon\right] \wedge \left[\mathbf{d}(\mathbf{Q}(E(x_i)), \mathbf{Q}(E(x_j)))\ge\delta\right]
\vspace{-4pt}
\label{eq:fair_def}
\end{equation}
where 1$\ge$$\epsilon$,$\delta$$>$0 
and $\mathbf{D}(\cdot)$, $\mathbf{d}(\cdot)$ are distance metrics applied to the input and output quantum spaces. 
\cite{guan2022verifying} has proved that if $\mathbf{D}(\cdot)$ and $\mathbf{d}(\cdot)$ are respectively specified as trace distance and measurement outcome distribution, then the following holds
\vspace{-4pt}
\begin{equation}
\underbrace{\mathbf{d}(\mathbf{Q}(E(x_i)), \mathbf{Q}(E(x_j)))}_\text{output measurement distance} \le \mathbf{K}\cdot\underbrace{\mathbf{D}(E(x_i), E(x_j))}_\text{input trace distance} 
\vspace{-6pt}
\label{eq:fair_k}
\end{equation}
where $\mathbf{K}$ (0$<$$\mathbf{K}$$\le${1}) is the Lipschitz constant that characterizes a  QNN circuit on a NISQ device. The smallest $\mathbf{K}$, denoted as $\mathbf{K^*}$, sets the upper limit on the ratio of $\delta$ to $\epsilon$ for a fair QNN. \textit{A smaller $\mathbf{K^*}$ indicates a more fair QNN}. A QNN is \textit{($\epsilon$, $\delta$)-fair} if and only if $\delta\ge\mathbf{K^*}\epsilon$.
Consider a noiseless QNN, $\mathbf{Q}$, and its noisy counterpart $\mathbf{Q_{Err}}$, which includes only depolarizing noise with an error rate of $p$. Let $\mathbf{K^*}$ and $\mathbf{K_{Err}^*}$ respectively represent the Lipschitz constants of $\mathbf{Q}$ and $\mathbf{Q_{Err}}$, we have 
\vspace{-6pt}
\begin{equation}
\mathbf{K_{Err}^*} = (1-p)\cdot{\mathbf{K^*}}
\vspace{-6pt}
\label{eq:fair_k_noise}
\end{equation}
Based on Equations~\ref{eq:fair_k} and \ref{eq:fair_k_noise}, we can utilize $p$ as a proxy for fairness if noises in a synthesized QNN circuit on a particular NISQ device can be unified as depolarizing noise. Note that $\bm{K^*}$ can be calculated based on $p$~\cite{guan2022verifying} to quantify the unfairness of a QNN. For simplicity, this work only presents the measured $p$ as a proxy for fairness and omits the $\bm{K^*}$ calculations.

\begin{figure}[t!]
\centering
\includegraphics[width=1\linewidth]{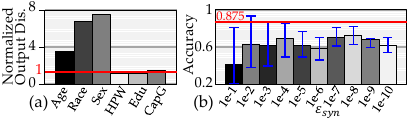}
\vspace{-24pt}
\caption{(a) Output distance and (b) accuracy of synthesized QNNs on \texttt{IBM\_Almaden} (HPW: hours-per-week, Edu: education; CapG: capital-gain).}
\label{f:motivation}
\vspace{-18pt}
\end{figure}

\textbf{Training \& Deployment}.
QNNs are implemented using multi-qubit gates and trained via hybrid quantum-classical methods. Optimized QNNs are then deployed onto NISQ devices, such as IBM computers, which support a specific native gate set consisting of two 1-qubit gates (\texttt{U2}, \texttt{U3}) and one 2-qubit gate (\texttt{CNOT}). Current 2-qubit gates, with an error rate of $10^{-3}$ and a coherence time of ten to a hundred microseconds~\cite{huang2019fidelity}, lead to considerable performance degradation.
To address this, quantum compilers~\cite{patel2022quest, younis2021berkeley} use approximate synthesis to reduce 2-qubit gate number (i.e., $N_{2q}$) and circuit depth in the synthesized QNN (i.e., $\mathbf{Q^{'}}$) while ensuring that the unitary difference between $\mathbf{Q^{'}}$ and the uncompiled $\mathbf{Q}$ remains within a predefined budget denoted as $\epsilon_{syn}$.
As shown in Figure~\ref{f:background}, approximate synthesis has three steps:
(1) The original $\mathbf{Q}$ is partitioned into blocks (i.e., $U_1$, $\cdots$, $U_N$) with each block having at most $S_{blk}$ (e.g., 3 ) qubits.
(2) The compiler searches for circuit candidates with a minimal $N_{2q}$ within an $\epsilon_{syn}$ budget over a tree~\cite{patel2022quest}. It generates a list of acceptable compiled circuits for each individual partition, e.g., \{$U_1^1$, $U_1^2$, $\cdots$, $U_1^{n_1}$\} for $U_1$. Note that the number of synthesized circuits can vary across different partitions.
(3) One compiled circuit is selected from the candidate list of each partition and sequentially recombined to form the synthesized QNN $\mathbf{Q^{'}}$.

\begin{figure*}[t]\centering
\center{\includegraphics[width=.98\textwidth]{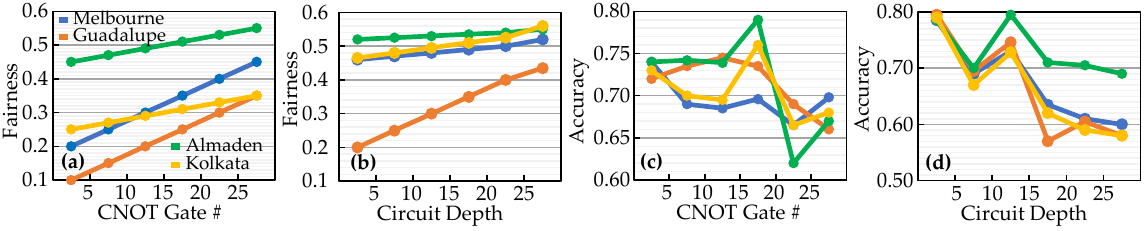}}
\vspace{-12pt}
\caption{QNN fairness vs. (a) \texttt{CNOT} gate \# and (b) circuit depth; QNN accuracy vs. (c) \texttt{CNOT} gate \# and (d) circuit depth.}
\label{f:design-space}
\vspace{-20pt}
\end{figure*}

\underline{\textbf{Limitations in Prior QNNs}}. 
We conducted a preliminary study to evaluate fairness/accuracy of the best QNN circuit generated by noise-aware training~\cite{wang2022vqcdac} using \textit{Adult Income}~\cite{adult_income} dataset. 
We compiled the circuit using default settings in BQSKit~\cite{patel2022quest, younis2021berkeley}. 
We then selected data pairs ($x_i$, $x_j$) with a unit input distance $\mathbf{D}(E(x_i), E(x_j))$, resulting from differences within a single feature group, e.g., \textit{age} or \textit{hour-per-week (HPW)}. 
We normalized the output distance to that of \textit{HPW} and presented the results in Figure~\ref{f:motivation}(a). \textit{The results reveal unfairness between different feature groups: the same unit difference in \textit{Sex} and \textit{HPW} results in a 7.8$\times$ difference in output distance.}
We also compile the QNN with various $\bm{\epsilon_{syn}}$ and report the averaged accuracy in Figure~\ref{f:motivation}(b) where the blue line denotes the maximal/minimal accuracy achieved for each $\bm{\epsilon_{syn}}$, while the red line is the expected ideal accuracy. \textit{Our observation reveals significant reductions and variations in accuracy after QNN synthesis, emphasizing the critical importance of QNN deployment over training in achieving model accuracy}. This is due to the potential compromise of accuracy gains from training by imprudent synthesis settings, which are often unknown to average end-users.

\vspace{-8pt}
\subsection{Reinforcement Learning}
\vspace{-4pt}
We use deep Q-learning (DQL)~\cite{hasselt2010double} that combines a lightweight reinforcement learning (RL) model with deep neural networks (DNNs) to search for the twin objectives of fairness and accuracy in QNN deployment, following its recent success in solving multi-constrained optimization problems in resource-limited scenarios~\cite{fan2020theoretical}.

\begin{wrapfigure}{b}{0.25\textwidth}
\begin{center}
\vspace{-6pt}
\includegraphics[width=0.26\textwidth]{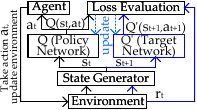}
\end{center}
\vspace{-16pt}
\caption{A standard DQL model.}
\label{f:RL_Overall}
\vspace{-10pt}
\end{wrapfigure}
\vspace{-12pt}
Figure~\ref{f:RL_Overall} shows a standard DQL model. A current state $s_t$ guides the selection of action $a_t$ from the action space $A(s_t)$ using a policy DNN. A target DNN, mirroring the policy DNN, aids in updating policy network parameters. The policy DNN takes $s_t$ as input and outputs action values for all possible actions, with $Q_{max}(s_t, a_t)$ indicating the best choice. Upon taking action $a_t$, the agent receives a reward $r_t$ and transitions to next state $s_{t+1}$. Accurate prediction of $Q(s_t, a_t)$ by the policy DNN satisfies Equation~\ref{eq:Policy_Target}. The loss function in Equation~\ref{eq:Loss} is employed for updating the parameters of both the policy and target DNNs.
\vspace{-4pt}
\begin{eqnarray}
\begin{aligned}
Q_{max}(s_t,a_t)=Q_{max}^{'}(s_{t+1},a_{t+1})+r_t
\end{aligned}
\label{eq:Policy_Target}
\end{eqnarray}
\vspace{-18pt}
\begin{eqnarray}
\begin{aligned}
Loss=[r+\gamma{\cdot}Q_{max}^{'}(s_{t+1},a_{t+1})-Q_{max}(s_t,a_t)]^2
\end{aligned}
\label{eq:Loss}
\end{eqnarray}

\noindent
\underline{\textbf{Challenges in RL-based QNN Deployment}}.
We are the first to apply DQL in QNN deployment. Our work identifies the following challenges and provides corresponding solutions.
\vspace{2pt}
\begin{enumerate}[leftmargin=*, topsep=0pt, partopsep=0pt, itemsep=0pt]
\item \textbf{States Specification}. 
The agent's state tensors must contain essential information to determine block connections that meet fairness and accuracy requirements. However, generating a comprehensive and high-quality state representation for QNN circuit deployment is a non-trivial task.

\item \textbf{Action Specification}.
Each agent action requires an exhaustive search in the complete action space for the current state. In addition, the action specification must align with the input/output structure of the policy/target DNN, ensuring a constant size for the state representation and action specification throughout the design process.

\item \textbf{Reward Quantification}. 
Balancing fairness and accuracy for different QNN designs is challenging, particularly when the QNN is only a partial circuit in the synthesis process.
\end{enumerate}

\section{Design Space Exploration}
\label{sec:exploration}
\vspace{-2pt}

\subsection{Design Space Complexity}
\vspace{-2pt}
We explore the design space complexity by performing approximate synthesis~\cite{patel2022quest, younis2021berkeley} using four state-of-the-arts QNNs~\cite{Sukin2019_vqcc14, chu2022qmlp, PatelST22vqcdate, wang2022vqcdac} onto \texttt{IBM\_Almaden}. 
We adjust the block size ${S_{blk}}$ and threshold $\epsilon_{syn}$ to control the number of generated synthesis blocks and the size of the search space. Increasing ${S_{blk}}$ and decreasing $\epsilon_{syn}$ result in a smaller search space. Table~\ref{tab:qnns_space} shows the results using settings of ${S_{blk}}$=2, $\epsilon_{syn}$=$10^{-2}$, and ${S_{blk}}$=3, $\epsilon_{syn}$=$10^{-5}$. It is worth mentioning that the results of DATE22~\cite{PatelST22vqcdate} deviate from the other three QNN circuits due to its unique circuit, using \texttt{CNOT} gate instead of parameterized entangled gates found in the other QNNs~\cite{Sukin2019_vqcc14, chu2022qmlp, wang2022vqcdac}. Overall, we observe significant differences in the search space for varying ${S_{blk}}$ and $\epsilon_{syn}$. While reducing the search space effectively reduces the search time, circuits obtained through a smaller search space generally perform worse than those obtained through a larger search space. 
\textit{Therefore, the vast complexity of the design space in practical QNN synthesis poses a significant challenge, requiring automated and efficient exploration strategies}.

\begin{table}[t]
\centering
\footnotesize
\caption{Comparison on Design Space Complexity.}
\vspace{-8pt}
\label{tab:qnns_space}
\begin{tabular}{|l|c|c|c|c|c|}\toprule
\multirow{2}{*}{\textbf{QNNs}}   &\multicolumn{2}{c|}{\textbf{Config.}}  & \multirow{2}{*}{\textbf{Space}} & \multirow{2}{*}{\textbf{Acc.}} & \multirow{2}{*}{\textbf{Fairness}} \\\cline{2-3}
& $S_{blk}$ & \textbf{$\epsilon_{syn}$}  & & & \\\midrule
\multirow{2}{*}{C14~\cite{Sukin2019_vqcc14}} &2 &1E-2 &43,046,721 &0.764 &0.3488 \\\cline{2-6}
& 3& 1E-5&729 & 0.724&0.4801\\\hline
\multirow{2}{*}{QMLP~\cite{chu2022qmlp}}  &2 &1E-2 &43,046,721 &0.772 &0.5126\\\cline{2-6}
& 3& 1E-5& 729&0.756 &0.4923\\\hline
\multirow{2}{*}{DATE22~\cite{PatelST22vqcdate}}  &2 &1E-2 &98,304 &0.624 &0.4992 \\\cline{2-6}
& 3& 1E-5&6,561 &0.764 &0.4697\\\hline
\multirow{2}{*}{DAC22~\cite{wang2022vqcdac}}  &2 &1E-2 &43,046,721 &0.748 &0.3293 \\\cline{2-6}
& 3& 1E-5& 6,561& 0.742&0.4985\\\bottomrule
\end{tabular}
\vspace{-20pt}
\end{table}

\begin{figure*}[t!]
\centering
\includegraphics[width=\textwidth]{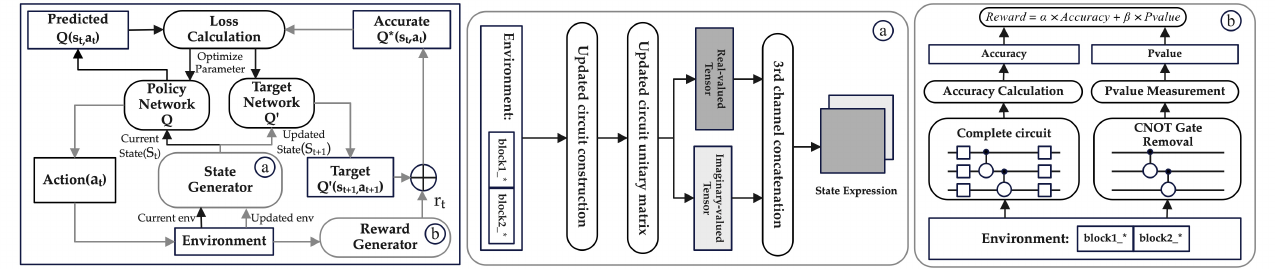}
\vspace{-0.26in}
\caption{The~\design framework. Left: the overall architecture. Middle and Right: functional blocks in~\design: (a) state generator; (b) reward generator.}
\label{f:RL_Design}
\vspace{-0.26in}
\end{figure*}

\vspace{-4pt}
\subsection{Fairness and Accuracy in Synthesized QNNs}
\vspace{-2pt}
We perform noise-aware training~\cite{wang2022vqcdac} and approximate synthesis~\cite{patel2022quest, younis2021berkeley} 
using a 8-qubit QNN~\cite{chu2022qmlp}. 
The target devices are \texttt{IBM\_Melbourne} and \texttt{IBM\_Guadalupe}.
To investigate the impact of \texttt{CNOT} numbers, we randomly selected a compiled circuit and iteratively reduced its \texttt{CNOT} count by one, while ensuring it remained within the $\epsilon_{syn}$ budget. Simultaneously, we recorded the corresponding fairness and accuracy measurements. We explore the impact of circuit depth caused by \texttt{CNOT} displacement by measuring fairness/accuracy for synthesized QNNs with varying depths. We repeated experiments 100 times and report the averaged results.


To evaluate QNN fairness, we utilize the measured noise error rate, $p$, on the target NISQ device for each synthesized circuit. Despite variations among NISQ computers, we observe a consistent monotonic increase in QNN fairness with higher \texttt{CNOT} count and circuit depth, as demonstrated in Figure~\ref{f:design-space}(a)-(b).
The QNN accuracy is intricately tied to both the \texttt{CNOT} count and circuit depth. Figure~\ref{f:design-space}(c) highlights that the optimal accuracy occurs at a middle sweet spot: exact synthesized circuits with 32 \texttt{CNOT} gates introduce error-prone 2-qubit gates, while drastically reducing \texttt{CNOT} gates (e.g., to 5) enlarges the unitary difference, both leading to a decrease in accuracy.
Figure~\ref{f:design-space}(d) demonstrates that, in most cases, reducing circuit depth enhances accuracy by operating within the qubit coherence time. However, interestingly, we observe a suboptimal spot across all IBM computers, with some instances (e.g., the 20-qubit \texttt{IBM\_Almaden}) achieving the best accuracy at certain circuit depths. 
\textit{In conclusion, existing compilers prioritize minimizing the \texttt{CNOT} gate count under an $\epsilon_{syn}$ budget. However, our findings demonstrate that the compiled QNN circuits exhibit substantial fairness and accuracy variations, as they are intricately correlated with both the \texttt{CNOT} gate number and circuit depth. These critical factors, previously overlooked by compilers, significantly influence the final performance of the compiled circuits.}

\section{\design}
\label{sec:design}
The relationship between QNN accuracy and fairness is complex and influenced by synthesized circuit characteristics, sometimes yielding contradictory results. The intricate design space and limited quantum circuit synthesis knowledge pose challenges for deploying fair and accurate QNNs on NISQ devices. To address these challenges, we propose \design, a DQL-based framework for automated fair and accurate QNN deployment. It includes a complete NISQ noise model and measurement method for precise noise assessment. Our framework also offers a flexible optimization objective with a customizable reward function, integrating fairness and accuracy to accommodate diverse application goals.

\vspace{-6pt}
\subsection{The~\design Framework}
\label{subsec:design_framework}
As illustrated in Figure~\ref{f:RL_Design}, the proposed~{\textit{\design}} framework primarily consists of a state generator, a reward generator, and paired policy/target networks. 
The policy/target networks are initialized using random parameters without prior training. The framework undergoes multiple exploration cycles. Each cycle starts from a blank design, which represents a completely disconnected synthesized QNN with multiple partitions. Throughout these cycles, the framework iteratively refines the blank design by taking actions, gradually shaping it into an optimized QNN configuration.

\textbf{Training Process}.
To initiate the exploration cycle, a random circuit from the first QNN partition, e.g., $U_1^i$, is selected as the initial action. Subsequently, the policy network produces a vector encompassing all possible actions that can be taken in that region, i.e., connecting $U_1^i$ with each candidate synthesized circuit for $U_2$. The overall reward is calculated, and this information, along with state, action, and value estimates, trains the neural network and updates the policy/target networks. The exploration cycle iterates to optimize the design. Once the search is complete, full system simulations are conducted to verify and evaluate the design.
In general, the policy network generates coarse designs, while the target network efficiently refines them based on prior knowledge, continuously generating more optimal configurations. Unlike traditional supervised learning, this framework doesn't require a training dataset; instead, the policy/target networks gradually train themselves from past exploration cycles.

\vspace{-8pt}
\subsection{The \design Error Model and Measurement}
\label{subsec:design_model}
\vspace{-4pt}

\textbf{Error Estimation}.
Prior research~\cite{chu2022qmlp, PatelST22vqcdate, wang2022vqcdac, guan2022verifying} focused on gate-level noise while neglecting \textit{crosstalk} errors. To address this, we introduced a noise-estimation circuit~\cite{urbanek2021mitigating} that integrates both \textit{gate} errors
and \textit{crosstalk} errors. Enhancing the method further, we incorporated randomized compiling~\cite{wallman2016noise} to convert incoherent errors to coherent errors, and quantum circuit engineering~\cite{ahsan2022quantum} to account for the impact of \texttt{CNOT} displacement on crosstalk errors.
The key steps of our approach are as follows:
(1) \textit{Estimation Circuit Construction}:
We start with a given QNN partition circuit, referred to as the target circuit. Then, we construct an error estimation circuit by removing all 1-qubit gates from the target circuit. This simplifies the circuit, making it easily simulatable and measurable, as shown in~\cite{urbanek2021mitigating}.
(2) \textit{Randomized Compiling and Circuit Engineering}:
We utilize randomized compiling~\cite{wallman2016noise} to convert coherent errors into incoherent errors, which involves inserting a layer of randomized 1-qubit gates before and after each layer of 2-qubit gates.
To incorporate the impact of \texttt{CNOT} displacement on crosstalk errors, we adopt the approach proposed in~\cite{ahsan2022quantum}. By varying the displacement of \texttt{CNOT} gates in each target circuit and creating multiple circuits for later measurement, we can effectively account for the impact of crosstalk errors and enhance the error modeling of our QNNs.

\textbf{Error Measurement}.
For each estimation circuit we synthesize, we measure its error rate on the target NISQ computer.
Note that we exclude readout errors from our error model because they are decoupled from the QNN circuits. However, we adopt the IBM calibration errors and mitigate their negative impact on QNN performance using previously proposed correction methods~\cite{nachman2020unfolding}.

\subsection{RL-based QNN Deployment}
\label{subsec:design_adaptation}

\textbf{Representation of QNNs (States)}.
In~\design, each state corresponds to a partial QNN circuit. The state tensor is constructed to accurately predict the characteristics of the final completed circuit. To achieve this, a natural choice is to utilize quantum unitary matrices for the partial circuit as they offer a complete representation of a quantum state. However, the quantum unitary involves complex values, making it unsuitable for direct processing by the policy/target network. Therefore, we propose a state generator denoted as {\large\ding{182}} in Figure~\ref{f:RL_Design}.
The key idea is to split the complex-valued QNN unitary matrix corresponding to the state into two parts: the real-valued tensor and the imaginary-valued tensor. These two tensors are then concatenated along the third dimension to form the input to the policy/target networks. The output tensor of the action decision network is sliced based on the current state's action space, and the index of the maximum value in this slice represents the action to be taken in the current state.

\textbf{Representation of Block Selection (Actions)}.
To meet the requirement of a constant-sized action specification throughout the design process with an invariable DNN structure, we sum up the candidate blocks from all partitions and use the combined total as the output size for both the policy and target networks. This approach guarantees that the size of the output tensor remains constant, irrespective of the number of blocks available for selection.
In this setting, the action space for each state is closely related to the block index corresponding to the current state. For instance, when the current state is $U_1^*$, the action space would be \{$U_2^1$, $U_2^2$, $\cdots$, $U_2^{n_2}$\}.

\textbf{Flexible Optimization Goal (Reward)}.
To ensure the integration of both fairness and accuracy in QNN deployment, a crucial aspect is designing a comprehensive reward function that guides the search effectively. To accommodate diverse design goals for various applications, we formulate a reward function that combines both fairness and accuracy while maintaining flexibility through the use of scaling factors, as shown in Equation~\ref{eq:reward_func}. 
To strike a balance between fairness and accuracy, QNN designers can adjust the values of ($\alpha$, $\beta$). For example, by setting ($\alpha$=0.5, $\beta$=0.5), a balanced QNN deployment is achieved, with equal emphasis on both fairness and accuracy. Adjusting the values of $\alpha$ and $\beta$, such as setting ($\alpha$=0.6, $\beta$=0.4) or ($\alpha$=0.4, $\beta$=0.6), we can prioritize either fairness or accuracy over the other metric.
\vspace{-4pt}
\begin{equation}
Reward = \alpha\cdot{Fairness} + \beta\cdot{Accuracy}
\label{eq:reward_func}
\vspace{-2pt}
\end{equation}

To maintain a predefined reward, we assess the accuracy and fairness of the partial QNN circuit after each action. As denoted as {\large\ding{183}} in Figure~\ref{f:RL_Design}, after each action, we measure the circuit's accuracy on real devices using the error model and workflow from the previous section. This estimation provides us with the error rate $p$, representing the fairness of the current QNN circuit. The final reward is then determined by combining both accuracy and fairness scores.

\begin{figure}[t!]
\centering
\includegraphics[width=1\linewidth]{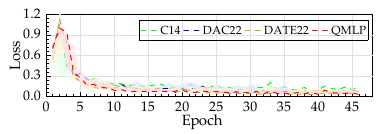}
\vspace{-24pt}
\caption{Loss changes during the training process.}
\label{f:RL_loss}
\vspace{-20pt}
\end{figure}

\begin{figure}[t!]
\centering
\includegraphics[width=1\linewidth]{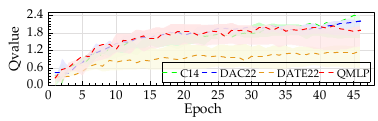}
\vspace{-24pt}
\caption{Q-value changes during the training process.}
\label{f:RL_Qvalue}
\vspace{-16pt}
\end{figure}

\begin{figure*}[t!]
\centering
\includegraphics[width=6.6in]{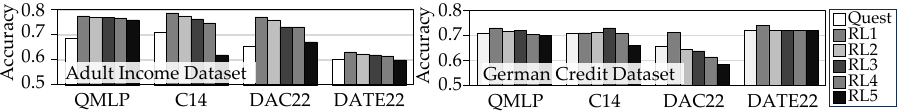}
\vspace{-0.16in}
\caption{Accuracy comparisons among different schemes.}
\label{f:Exp_Acc}
\vspace{-0.19in}
\end{figure*}

\begin{figure*}[t!]
\centering
\includegraphics[width=6.6in]{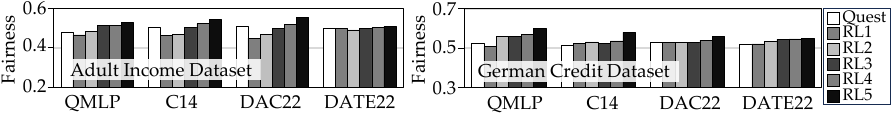}
\vspace{-0.16in}
\caption{Fairness scores comparisons among different schemes.}
\label{f:Exp_Fairness}
\vspace{-0.19in}
\end{figure*}
\begin{figure*}[t!]
\centering
\includegraphics[width=6.6in]{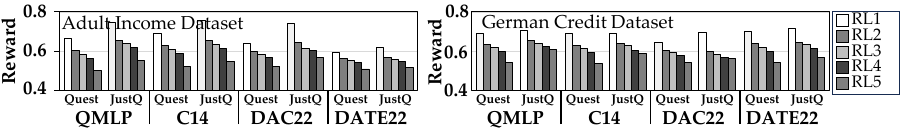}
\vspace{-0.16in}
\caption{Reward score comparisons among different schemes.}
\label{f:Exp_acc_fair}
\vspace{-0.29in}
\end{figure*}

\section{Experimental Methodology}
\label{sec:method}

\noindent
\textbf{Datasets \& Benchmarks}. 
We use two multi-attribute financial datasets: \textit{Adult Income}~\cite{adult_income} and \textit{German Credit}~\cite{german_credit}, similar to~\cite{guan2022verifying}.
\textit{German Credit} contains 20 attributes, but we selected 8 based on their influence on credit prediction. We obtained a final dataset of 1000 loan applicants, split into 700 for training and 300 for testing due to quantum resource limitations. For \textit{Adult Income}, we used a modified version with 8 attributes, randomly sampling 800 data points for training and 300 for testing. We evaluate~\design using four recent QNNs: C14~\cite{Sukin2019_vqcc14}, QMLP~\cite{chu2022qmlp}, DATE22~\cite{PatelST22vqcdate}, and DAC22~\cite{wang2022vqcdac}. 
C14 is a VQC designed for classification. QMLP utilizes a multilayer perceptron architecture. Both DATE22 and DAC22 employ adaptive circuits to enhance their performance.

\noindent
\textbf{Synthesis and NISQ Computers}.
We use BQSKit~\cite{patel2022quest, younis2021berkeley} for approximate synthesis and Qiskit to deploy the synthesized QNNs on NISQ devices. 
We run experiments on six NISQ computers: 
14-qubit \texttt{IBM\_Melbourne},
16-qubit \texttt{IBM\_Guadalupe},
20-qubit \texttt{IBM\_Almaden},
20-qubit \texttt{IBM\_Boeblingen},
27-qubit \texttt{IBM\_Auckland},
and
27-qubit \texttt{IBM\_Kolkata}.

\noindent
\textbf{Reinforcement Learning}. 
We use DQL with ResNet19 as the policy/target network backbone. 
The search process comprises 1000 iterations to determine the optimal circuit. 
During RL training, we set the learning rate and decay factor $\gamma$ respectively as 1e-3 and 0.99. We gradually decrease the exploration probability for $\epsilon$-greedy from 0.05 to a final value of 1e-2.

\noindent
\textbf{Schemes}.
We use the default approximate synthesis~\cite{patel2022quest, younis2021berkeley}, denoted as \textit{Quest}, as our baseline.
We evaluate five different~\design configurations against \textit{Quest}:
(1) \textit{RL1}: $\alpha$=0.1, $\beta$=0.9;
(2) \textit{RL2}: $\alpha$=0.4, $\beta$=0.5;
(3) \textit{RL3}: $\alpha$=0.5, $\beta$=0.5;
(4) \textit{RL4}: $\alpha$=0.6, $\beta$ = 0.4;
(5) \textit{RL5}: $\alpha$=0.9, $\beta$=0.1.

\section{Results and Analysis}
\label{sec:results}

\noindent
\textbf{Trainability of~\design}.
Figures~\ref{f:RL_loss} and \ref{f:RL_Qvalue} visualize the training process, showing a consistent decrease in the loss and a steady increase in the Q-value. For instance, in the case of QMLP, the initial training loss value of 0.5356 decreased to 0.0865 by the 20th iteration, while the Q-value increased from 0.31 to 1.8382. Beyond this point, both the loss and Q-value stabilize without significant further changes. These results demonstrate the effective trainability of the \textit{\design} model for nontrivial QNN deployment tasks.

\noindent
\textbf{Performance Results}.
We compared the performance of synthesized circuits generated by the baseline \textit{Quest} and \textit{\design} on all QNN benchmarks used in this study. 
Figure~\ref{f:Exp_Acc} and Figure~\ref{f:Exp_Fairness} demonstrate that circuits generated by \textit{\design} outperform those from the baseline \textit{Quest} compiler in terms of accuracy and fairness metrics when appropriate weights ($>$0.5) are applied.
We also assess the trade-off between accuracy and fairness by comparing the weighted results of the circuits, as shown in Figure~\ref{f:Exp_acc_fair}. The \textit{\design} output circuit achieves a higher weighted sum of accuracy and fairness compared to the \textit{Quest} output circuit.
In summary, \textit{Quest} underperforms due to its dual-annealing-based minimization approach~\cite{patel2022quest}, which prioritizes output distance over fairness and accuracy, leading to imbalanced circuits. In contrast, \textit{\design} achieves a superior balance between accuracy and fairness in the synthesized circuits.

The scaling weights in Equation~\ref{eq:reward_func} are crucial for balancing accuracy and fairness. Results in Figure~\ref{f:Exp_acc_fair} show that higher accuracy weight leads to higher accuracy but lower fairness in the output circuit.
For instance, DAC22 generated by \textit{Quest} achieved an accuracy of 0.656 and a fairness score of 0.5101 on \textit{Adult Income}. When balancing accuracy and fairness with ($\alpha$=0.5, $\beta$=0.5), the resulting overall score is 0.583. However, the \design approach, under the same scaling weights, achieved an accuracy of 0.732 and a fairness score of 0.5546, resulting in a higher overall score of 0.6433. Notably, the \design output circuit outperformed the circuits generated by \textit{Quest} algorithm on all three metrics (i.e., accuracy, fairness, and weighted result), highlighting the potential benefits of this approach.

\vspace{-2pt}
\section{Conclusion}
\vspace{-4pt}
This work initiates research on fair and accurate QNNs by exploring the design space. We emphasize the deployment phase's critical role and its influence on QNN accuracy and fairness. We propose JustQ, a reinforcement learning-based framework for QNN deployment. Experimental results demonstrate JustQ's superiority, producing synthesized QNN models that excel in accuracy and fairness. 

\vspace{-2pt}
\section*{Acknowledgments}
\vspace{-2pt}
This work was supported in part by NSF CAREER AWARD CNS-2143120.
Any opinions, findings, and conclusions or recommendations expressed in this material are those of the authors and do not necessarily reflect the views of grant agencies or their contractors.

\vspace{-4pt}
\footnotesize
\bibliographystyle{ieeetr}
\bibliography{main-aspdac.bib}

\end{document}